\documentclass[fonts]{icst}

\pdfoutput=1

\usepackage{moreverb}

\usepackage[breaklinks,colorlinks,bookmarksopen,bookmarksnumbered,linkcolor=ICSTblue,citecolor=ICSTblue,urlcolor=ICSTblue]{hyperref}
\usepackage{breakurl}
\usepackage{doi}

\newcommand\BibTeX{{\rmfamily B\kern-.05em \textsc{i\kern-.025em b}\kern-.08em
T\kern-.1667em\lower.7ex\hbox{E}\kern-.125emX}}

\def\volumeyear{2012}

\begin{document}

\runningheads{M. Convertino, et al.}{Power-law of Aggregate-size Spectra}
%%%%%%%

% \title{\textbf{From River Basins to Elephants to Bacteria Colonies: a Universal Model of Aggregate Organization for Animate and Inanimate species}}

%\title{\textbf{From River Basins to Elephants to Bacteria Colonies: Aggregate Allometry of Animate and Inanimate species}}

%\title{\textbf{From River Basins to Elephants to Bacteria Colonies: aggregate-size Spectrum of Animate and Inanimate Species}}

%%%
%%%
%%%

%\title{\textbf{On the Power-law of the aggregate-size Spectrum in Natural Systems}}

\title{Power-law of Aggregate-size Spectra in Natural Systems}

\author{Matteo Convertino\affil{1}\affil{2}\affil{3}\fnoteref{1}, Filippo Simini\affil{4}, Filippo Catani\affil{5}, Igor Linkov\affil{2}\affil{6}, Gregory A. Kiker\affil{1}\affil{3}}

\address{\affilnum{1}Department of Agricultural and Biological Engineering - IFAS, University of Florida, Gainesville, FL, USA\\
\affilnum{2}Florida Climate Institute, c/o University of Florida, and Sustainable-UF, Gainesville, FL, USA\\
\affilnum{3}Contractor of the US Army Corps of Engineers at the Risk and Decision Science Team, Engineer Research and Development Center (ERDC), Concord, MA, USA\\
\affilnum{4}Center for Complex Network Research (Barabasi Lab), Department of Physics, Northeastern University, Boston, MA\\
\affilnum{5}Engineering Geology and Geomorphology Unit, Department of Earth Sciences, Universit\`{a} di Firenze, Firenze, IT\\
\affilnum{6}Department of Engineering and Public Policy, Carnegie Mellon University, Pittsburgh, PA, USA}

%\author{\textbf{M. Convertino$^{a,b,c \; 1}$, F. Simini$^{d}$, F. Catani$^{e}$, I. Linkov$^{b,f}$, G.A. Kiker$^{a,c}$}}

%%%% VERY SHORT ABSTRACT 
%%%%%%Patterns of animate and inanimate systems show remarkable similarities in their aggregation. One similarity is the double-Pareto distribution of the aggregate-size of system components. Not many models have been developed to describe probabilistically the aggregate-size distribution of any system. Considering aggregates as islands and their perimeter as a curve mirroring the sculpting network of the system, the probability of exceedence of the drainage area, and the HackÕs law are shown to be the the Kor?akÕs law and the perimeter-area relationship for river basins. The perimeter-area relationship, and the probability of exceedence of the aggregate-size provide a meaningful estimate of the same fractal dimension. The aggregate-size distribution is accurately derived using the null-method of box-counting on the occurrences of system components. The importance of the aggregate-size spectrum relies on its ability to reveal system form, function, and dynamics.

\abstract{Patterns of animate and inanimate systems show remarkable similarities in their aggregation. One similarity is the double-Pareto distribution of the aggregate-size of system components. Different models have been developed to predict aggregates of system components. However, not many models have been developed to describe probabilistically the aggregate-size distribution of any system regardless of the intrinsic and extrinsic drivers of the aggregation process. 
Here we consider natural animate systems, from one of the greatest mammals - the African elephant (\textit{Loxodonta africana}) - to the \textit{Escherichia coli} bacteria, and natural inanimate systems in river basins. Considering aggregates as islands and their perimeter as a curve mirroring the sculpting network of the system, the probability of exceedence of the drainage area, and the Hack's law are shown to be the the Kor\v{c}ak's law and the perimeter-area relationship for river basins. The perimeter-area relationship, and the probability of exceedence of the aggregate-size provide a meaningful estimate of the same fractal dimension. Systems aggregate because of the influence exerted by a physical or processes network within the system domain. The aggregate-size distribution is accurately derived using the null-method of box-counting on the occurrences of system components. 
The importance of the aggregate-size spectrum relies on its ability to reveal system form, function, and dynamics also as a function of other coupled systems. Variations of the fractal dimension and of the aggregate-size distribution are related to changes of systems that are meaningful to monitor because potentially critical for these systems.} %Hence, the importance to accurately assess the aggregate-size spectrum. }


%\tnotetext[1]{Please ensure that you use the most up to date
%class file, available from ICST at \url{http://icst.org/icst-transactions/}
%}

\fnotetext[1]{Corresponding author.  Email: \email{mconvertino@ufl.edu}}

\maketitle

%\title{\textbf{Power-law of Aggregate-size Spectra in Natural Systems}}
%
%
%
%\author{\textbf{M. Convertino$^{a,b,c \; 1}$, F. Simini$^{d}$, F. Catani$^{e}$, I. Linkov$^{b,f}$, G.A. Kiker$^{a,c}$}}
%\vspace{1cm}
%
%\affil{$^{a}$ Department of Agricultural and Biological Engineering-IFAS, University of Florida, Gainesville, FL, USA}
%\affil{$^{b}$ Risk and Decision Science Team, Engineer Research and Development Center (ERDC), US Army Corps of Engineers, Concord, MA, USA} 
%\affil{$^{c}$ Florida Climate Institute, c/o University of Florida, Gainesville, FL, USA}
%\affil{$^{d}$ Center for Complex Network Research (Barabasi Lab), Department of Physics, Northeastern University, Boston, MA}
%\affil{$^{e}$ Engineering Geology and Geomorphology Unit, Department of Earth Sciences, Universit\`{a} di Firenze, Firenze, IT}
%\affil{$^{f}$ Department of Engineering and Public Policy, Carnegie Mellon University, Pittsburgh, PA, USA}

%\maketitle
%\thispagestyle{empty}
%
%
%$^1$ {\it Corresponding author:} M. Convertino, Department of Agricultural and Biological Engineering-IFAS, Frazier Rogers Hall, Museum Road, PO box 110570, 32611-0570; and, Risk and Decision Science Team, Engineering Research \& Development Center, US Army Corps of Engineers, New England District, Concord, 01742, MA (email: mconvertino@ufl.edu; phone: +1 781-645-6070; fax: +1 352-392-4092).\\

%{\it Keywords: aggregate-size, fractal dimension, river basins, networks, systems, allometry}\\

\section{Introduction}

The understanding of the causes underlying the spatial organization of species in ecosystems is one of the most challenging and debated topics in ecology. This is also true for the spatial organization of components of other systems, such as inanimate natural systems. This is for example the case of river basins. As for human-made systems, the assemblage of these systems is mostly determined by human design; however, the human component dynamics makes these systems not completely deterministic. This is for example the case of cities.
Questions arise about the level of complexity of theories and models to reproduce and characterize the aggregation of systems components. Here a \textit{system component} is broadly defined as the elementary unit that forms animate (biotic) or inanimate (abiotic) aggregates indistinctly. For example individuals of the same species form aggregates in an ecosystems and the whole metapopulation is defined by the whole set of aggregates.
In living or animate systems, aggregates of systems components are observed from the microscale (for example, bacteria \cite{rinaldosize02,buldyrev03}), the meso/macroscale (for example, cancer cells \cite{marco09}, and ants \cite{vandermeer08}), to the continental scale (for example, trees \cite{rex90,Sole95,plotkin00,simini10,conv11ecomod}, fishes \cite{conv11ecomod}, African elephant \cite{Loarie2009}, and corals \cite{Purkis07}). In non-living or inanimate systems aggregates are observed as well at different scales \cite{Stanley96}. 
Concepts developed for animate system components were generalized to inanimate system components. For instance, as in \cite{Banavar00,maritan02}, and in \cite{ignacio11}, river basins can be considered as living systems if we consider them as animate systems (organisms) characterizable by a metabolism (proportional to the evapotranspiration and the drainage area) and a body-mass (proportional to the area of connected tributaries). Each subbasin is formed by canal pixels and hillslope pixels that are both system components. Analogies to organisms have also been formulated for cities \cite{Samaniego08}. This analogy between animate and inanimate systems allows a potentially mutual understanding of systems, and the ability to adopt similar probabilistic methods to characterize both systems. It is certainly difficult to claim universal principles of organization of systems; however, the development of methods to characterize both animate and inanimate systems is certainly useful for monitoring these systems and for system design.

One of the most fundamental variable characterizing animate and inanimate systems is the aggregate-size.
The aggregate-size (named as patch-size, or cluster-size in ecology) is defined as the area of the landscape in which individuals of systems are aggregated together \cite{Kuninbook,Kunin2003}. In ecological modeling the definition of aggregates for species is generally a non-trivial task that requires the definition of many biological variables. These local system variables are the occurrence of species, the minimum area to support a population, the habitat quality, and the sex-structure of species to list just a few. The aggregates of species are generally the input of metapopulation models for determining species abundance by considering the stochastic dynamics of birth-death and dispersal in and among metapopulation aggregates. The stochastic dispersal occurs on a network that connects species aggregates. The occurrence of system components is certainly one of the most important variables in both defining the aggregate-size and for the inference of system dynamics. For example, in ecosystems the location of species occurrences is also useful to estimate the abundance of species \cite{hegaston07}, the relationships between species and environmental variables for the definition of niches and climate change effects on species \cite{elith09c}, and the interactions with other species \cite{azaele10}.

The importance of the aggregate-size relies also on its distribution within the system analyzed and the variation of this distribution in time. The probabilistic structure, and more precisely the distribution of the aggregate-size of systems components is widely reported to be a power-law. This is particularly the case for single and multiple species considered together. However, exponential probability distributions of the aggregate-size are observed for some species \cite{bonab99,niwa03,Hardenberg2010}, for perturbed and evolving ecosystems (for example, for vegetation due to grazing \cite{kefi07,kefi11}, and for ecosystems characterized by strong gradients of some environmental variables (for example, for vegetation in the Kalahari rainfall transect in Africa \cite{scanlonnature}). Power-laws of the aggregate-size for inanimate systems (natural and man-made), such as river-basins, landslides, snow-cover in landscapes \cite{shook93}, and cities \cite{Giesen10} are also observed. Even for inanimate systems deviations from the power-law distribution are observed: for example exponential aggregate-size distributions are reported for cities subjected to rapid urbanization \cite{Benguigui2010}), and log-normal distributions are observed for submarine landslides \cite{Brink09}. A consistent part of the literature investigates the origin of the power-law distribution of aggregates and the causes of deviations from this distribution \cite{kefi11}. Here we confine our interest in animate and inanimate natural systems whose distribution of the aggregate-size is a power-law which seems to occur in the majority of cases in which ecosystems are at stationary state in their evolution \cite{banavarriv01}, or around a stable state in the energy landscape \cite{kefi11}. One of the explanations for the power-law distribution of the aggregate-size provided by literature is related to the typology of species movement in ecosystems. Individuals of species, from bacteria to elephants, seem to follow a simple Brownian \cite{cedric11}, or a Brownian-L\'{e}vy movement \cite{Benhamou07,dejager12,Jansen12}. This typology of movement arises from an optimal foraging strategy determined for instance by an optimization of the species-dispersal for survival \cite{bonab99,getz08,Reynolds10,Muneepeerakul11} constrained by the environment topology and constraints (for example, the river network and basin ridges in a river basin, or a Petri dish for in vitro bacteria populations). This type of dispersal is generally simulated with a combination of exponential and ``heavy-tailed'' dispersal kernel \cite{Muneepeerakul11} that results in a scale-free distribution of the aggregate patches of the species simulated \cite{boyer06,conv11ecomod}.

Aggregates of species are linked together by physical networks (for instance, river networks) \cite{conv11ecomod} or process-networks \cite{Souza07} (for example, communication networks and dispersal networks that describe the communication and movement of individuals respectively) from which the observed patterns arise. Process-networks can be embedded into physical networks (for example, dispersal networks of fishes are constrained within the river network \cite{muneefish}) or can exist without a visible physical network in the system space (for instance, the communication network among bacteria colonies or among transceivers considering animate and inanimate systems respectively). Many detailed processes are responsible for the formation of aggregates, and many models were developed in literature to predict aggregates of species. Some models tried to mimic the fine-level details of ecological processes, such as species interactions (for example, conspecific attractions \cite{Fletcher09}) and feedbacks (for example, density-dependence) among species. Other non physical-based models were built around other ``macroscopic'' theories that consider the ensemble average behavior of systems components, such as the theory of self-organized criticality \cite{bak87,bak88,Satorras01,sornette06}, allelomimesis \cite{Juanico05}, preferential attachment \cite{Souza07}, metabolic optimization \cite{West99,Enquist09}, percolation \cite{He03,Gardner07}, habitat suitability, and the neutral theory of biodiversity \cite{hubbellbook,muneefish,Manor08,konar10,conv11ecomod}. Network-based model were developed on these theories to reproduce patterns of aggregation of complex systems. The network framework is a simplified and valuable framework that allows to capture average properties of systems dynamics and organization. An example is the theory of optimal channel networks (OCNs) \cite{fracriv} that without the inclusion of geomorphological details is capable to describe analytically the topological properties of river networks. Aggregation phenomena of species were successfully modeled using the framework of OCNs, or other network-based models.
While network-based model were developed to reproduce animate and inanimate processes, not many network-based models were developed to probabilistically describe patterns of these processes created using data or model predictions. At the same time no consistent advancement occurred in the development of analytical forms for the probability distribution of the aggregate-size. In literature different analytical forms are found for different types of power-law distribution of the aggregate-size.
The purpose of this study is (i) to provide insights into a parsimonious model (box-counting \cite{Liebovitch89}) for assessing aggregates of systems just using occurrences of systems components, (ii) to integrate theories of aggregate-size distributions ( the Kor\v{c}ak's law \cite{Korcak40}, the perimeter-area relationship, and the theory of fractal river basins \cite{fracriv}) for animate and inanimate species and test the validity of this integration on all the systems analyzed, and (iii) to formulate a generalized analytical form for all the types of power-law probability distributions of the aggregate-size. We particularly focus on systems that exhibit a power-law spectrum of the aggregate-size.

%% complex models, simple network based process models 
%%% not many models have been developed to characterize probabilistically the aggregate
%%% network based models for characterizing a system rather than for modeling the system 
%% talk about the lack of an analytical pdf that characterize power laws 

The paper is organized as follows. Section \ref{matmet} describes the assumptions, the theoretical framework, the data, and the models used to predict the aggregates of systems components. Section \ref{res} reports the results of the box-counting method, of other models, and the validation of the theory. The discussion of the results is in Section \ref{discussion} Section \ref{concl} lists the most important conclusions, perspectives for future research, and potential applications of our findings.

\section{Materials and Methods} \label{matmet}

\subsection{Kor\v{c}ak's law and Aggregation Hypothesis}
%% change landscape to ecosystems 

Studies about the prediction of aggregates of systems and their theoretical characterization was developed separately among scientific disciplines; this is because aggregation phenomena occur in a broad variety of systems of different nature. In geography it is well known that the size of islands follows a power-law probability distribution, $P(S \geq s) \sim s^{-b}$, in which the exponent of the exceedence probability distribution is related to the fractal dimension of the coast of the islands \cite{Korcak40}. This power-law probability is called ``Kor\v{c}ak's law''. The exponent is half of the fractal dimension of the island coastline ($b=1/2 \; D$) \cite{Mandelbrot82}. Mandelbrot \cite{Mandelbrot82} found that the average value of this exponent is 0.65, with variations from 0.5 for African islands to 0.75 for Indonesian islands; thus, $b$ is within the range $[1; 1.5]$.

Landscape ecology is the field in which the theory of aggregates received the highest attention due the importance of the species aggregate-size distribution for species conservation. \cite{nikora99} and \cite{Shankar02}, studied the patches of river ecosystem properties (for instance, slope, hydrogeology, erosion, and vegetation) in
New Zealand. These are the first studies, to the best of our knowledge, that
tried to unify theories, including the ``Kor\v{c}ak's law'', for the characterization
of patches in heterogeneous ecosystems. However, the fractal exponents derived in
these studies were considered as independent estimates of features of individual aggregates
and of the whole mosaic of aggregates. Moreover, these studies did not correlate any network of the ecosystem analyzed to the pattern of aggregates of system components.

In geomorphology the aggregation of subbasins around river networks was elegantly investigated by many studies, starting from \cite{Howard90}, to the comprehensive review of \cite{fracriv} in which the theory of optimal channel networks was proposed. In ecology, \cite{conv11ecomod} analyzed the aggregation of species in river networks and 2-D landscapes, considering the exponent of the exceedence probability distribution of the aggregate-size as a meaningful indicator of the collective organization of species. That exponent was considered as a function of geometrical and environmental constraints of the ecosystem where aggregates form.

In this paper the following hypotheses have been tested.

\begin{enumerate}

\item It is possible to predict the aggregates of animate and inanimate systems and the aggregate-size spectrum solely from the occurrences of systems components. In the case of inanimate systems we consider the center of mass of each aggregate as an occurrence of system component. The box-counting method on the occurrences of systems components is a reliable method for calculating the aggregate-size. The accuracy of the box-counting in the aggregate and aggregate-size predictions is assessed with respect to other methods based on prediction of aggregates' area and perimeter. 
% value of occurrences just those to calculate aggregates. no need of anything else. we use the box counting 

% In Figure \ref{fig1} we synthesize the main hypothesis of this paper. 

%We imagine aggregates as islands and systems networks (such river networks and dispersal networks) as coastlines. Aggregates and aggregates' boundaries are sculpted by network processes. We define $S$, $c$, and $L_{\|}$, as the size, perimeter, and diameter of aggregates respectively, and $l$ as the length of the aggregates' sculpting network. 

%These quantities are used in the theoretical construct explained in Section \ref{theory}. 

%This idealization derived from \cite{conv2007} can be applied also to subbasins which are aggregates of river basins. Considering this analogy we use the theory of fractal river basins and we test it against the Kor\v{c}ak's law and the perimeter-area relationship for assessing the fractal dimension of aggregate patterns. In any landscape a self-affine or self-similar network can be traced and statistical properties of aggregates can be sampled along the network such as in \cite{conv2007} for subbasins. 

\item The theory of fractal river networks can be generalized in order to characterize the probabilistic structure of the aggregate-size of other systems arranged along river networks (for example, landslides as inanimate systems, and fishes and trees as animate systems), and arranged along non-visible networks of system processes occurring in landscapes (for example dispersal networks). In any system a meaningful self-affine or self-similar network can be traced and statistical properties of aggregates can be sampled along the network such as in \cite{conv2007} for the subbasin drainage area. In fact, the coalescence of system components can be described as an aggregation phenomena along branching trees \cite{Gabrielov99,dacosta02}.
We consider the Kor\v{c}ak's law \cite{Korcak40}, that is the power-law probability distribution for the size of islands as the generic probability distribution for the aggregate-size of any animate and inanimate systems components. Thus, aggregates (such as subbasins and species aggregates) are considered as islands and systems networks (such as river networks and dispersal networks) as coastlines in analogy. Aggregates and aggregates' boundaries are sculpted by network processes. $S$, $c$, and $L_{\|}$ are defined as the size, perimeter, and diameter of aggregates respectively, and $l$ as the length of the aggregates' sculpting network (Figure \ref{fig1}).
We tested the analogy between the theory of fractal river basins against the Kor\v{c}ak's law and the perimeter-area relationship by assessing the fractal dimension of aggregate patterns for these three conceptual models. We believe about the existence of a unique fractal dimension estimated by these models. This fractal dimension is a representative indicator for the whole set of systems aggregates and of each aggregate on average.

%This idealization derived from \cite{conv2007} can be applied also to subbasins which are aggregates of river basins. 

%% theoretical model based on a network principle to characterize aggregates. we consider korcak that was not developed considering a network approach . but indeed starting from the knowledge of ocn theory we noticed some similarity. in fact islands are created along a network that is a submarine ridge. So the submarine ridge  

\item A novel analytical formulation for the double-Pareto probability distribution (or spectrum) of the aggregate-size of systems components is formulated. Such distribution can fit the Pareto distributions that animate and inanimate systems components show. 

% importance considering the availability of data 

\end{enumerate}

\subsection{Systems Data}

Animate and inanimate natural systems are considered for a wide range of body-mass of system components, climatological condition, and biological dynamics in order to verify the validity of the proposed probabilistic description. Available data of our current and past research studies allow to consider animate and inanimate systems at different spatial scales (Figure \ref{fig2}) that exhibit a power-law probability distribution of the aggregate-size of system components. These systems are the E. coli bacteria in nutrient-rich substrate (courtesy of \cite{buldyrev03}), the subbasins of a portion of the Tanaro basin (Italy) \cite{conv2007}, the Snowy Plover in Florida in 2006 \cite{convertinoNATO}, the historical landslides of the Arno river basin (Italy) \cite{catani05}, the African elephant in the Kruger National Park (KNP) in 2006 (South Africa) \cite{krugerdata}, and fish and tree species in the Mississippi-Missouri River System (MMRS) (USA) \cite{muneefish,konar10,conv11ecomod}. As for the trees of the MMRS we consider only big trees for which the diameter at breast height is larger or equal than five inches \cite{conv11ecomod}.

Figure \ref{fig2} shows the occurrences of the aforementioned systems in order of the extension of the system domain where they occur. The extension of these systems covers fifteen orders of magnitude from the Petri dish of the E. coli ($6.1 \times 10^{-9} \; km^2$), the Tanaro basin ($5.3 \times 10^2 \; km^2$), the beach habitat along the Gulf coast of Florida ($\sim 5.6 \times 10^2 \; km^2$), the Arno basin ($8.23 \times \; 10^3 km^2$), the KNP ($19.0 \times \; 10^3 km^2$), to the MMRS ($2.98 \times \; 10^6 km^2$). For the African elephant, the Snowy Plover, and the E. coli we evaluate one pattern of occurrences as a realization of a process in which aggregation always occurs \cite{convertino10scalesbirds,convele}. 
For the Snowy Plover occurrences are available from 2002 to 2011 obtained by field survey \cite{convertinoNATO}, and for the African elephant from 1985 to 2004 for the dry season obtained by plane survey. We anticipate that a temporal analysis of the occurrence patterns is the subject of forthcoming papers. Here we examine the years for which the reliability of the occurrence patterns is the highest, in terms of data quantity and data quality. Other yearly-sampled occurrences of both African elephant and Snowy Plover show the formation of very similar aggregates. This is the case also of the E. coli bacteria in which self-similar patterns are observed in Petri dishes \cite{Budrene95} for different values of the nutrient concentration.

\subsection{Box-counting}\label{boxc}

The first step of the box-counting method is the creation of a coarse grid of boxes to overlay on the top of the system domain analyzed. The grid is then refined at each step until the lower cutoff of the analysis. The box-counting technique \cite{Liebovitch89} leads to a scaling relationship between the number of boxes ($N(r)$) in which at least an occurrence of system components is contained and the length of the side of the box ($r$). The relationship is a power-law, $N(r) \sim r^{-D_b}$, where $D_b$ is the Minkowski-Bouligand dimension that is a good estimate of the fractal dimension (or Hausdorff dimension) of the point-pattern of occurrences analyzed. The box-counting technique is applied to all the point-patterns of Figure \ref{fig2} for at least $2^{16}$ orders of magnitude of $r$. The box-counting is illustrated in Figure \ref{fig2} (c) for the Snowy Plover occurrences. 
Variabilities of measured exponents ($D_b$) for different systems are expressed as standard errors found by a Maximum Likelihood Estimation method (Section red{mle}) bootstrapping over cases and deriving exponents using the linear and the jackknife models \cite{Warton06}. For the river basin, the landslides, and the E. coli colony patterns the center of mass of each system component (i.e. a subbasin, a landslide, and a E. coli colony respectively) is considered in the box-counting analysis. In Figure \ref{fig1} the center of masses of ideal aggregates are shown as grey dots. For the system with occurrence data of system components available (i.e., big trees and fishes of the MMRS, African elephant in the KNP, and SP along the Gulf coast of Florida), the point-patterns of occurrences are directly analyzed without any pre-processing. \\

\subsection{Models of Aggregate Prediction} \label{models}

Aggregates of systems considered in this paper are predicted by models based on different assumptions, hypothesis, and at different levels of complexity. In the following we give a brief explanation of the models. We remind the reader to papers in which each model was implemented for more details. The area of an aggregate is defined as the sum of adjacent pixels considering the Von Neumann neighboring criteria. The perimeter of an aggregate is defined by the sum of the sides of the external pixels composing the aggregate.

For river basins, landslides, and E. coli colonies the aggregates are extrapolated by an image analysis model. The observed E. coli pattern (courtesy of \cite{buldyrev03}) (Figure \ref{fig2}, a) is binarized by extracting pixels whose grayscale value is higher than 30 (white pixels are logical ``true''). This threshold allows to reproduce the observed patterns with an accuracy of 92 \%. The area and perimeter are calculated for all the aggregates extracted using the grayscale threshold criteria. The code for extracting and calculating the aggregates is developed by the first author using Matlab \cite{matlab}. 

The subbasins of the Tanaro basin in Figure \ref{fig2} (b) are derived in \cite{conv2007} by extracting the river network from the digital elevation model (DEM). The network extraction is based on the identification of the contributing areas for each stream of the network. The extraction of the network and other hydrogeomorphological analysis are performed using the free software HydroloGis \cite{hydrologis}. As for the landslides in the Arno basin the over 27,500 recorded landslide occurrences were identified in \cite{catani05} using aerial-photo interpretation, expert knowledge, and remote sensing techniques.  Details are explained in \cite{catani05}.

For the SP (Figure \ref{fig3}, b) a habitat suitability model coupled with a patch delineation model is used to determine the aggregates of species \cite{convertino10time}. \cite{convertino10time} defined as a shorebird aggregate an aggregate of pixels whose habitat suitability index is higher than a certain threshold, big enough to support all together a meaningful population size but not too small to support at least a breeding pair, and close enough to support breeding and wintering activity. The habitat suitability index is based on habitat suitability maps predicted by a maximum entropy model \cite{phillips06,phillips08b} constrained on environmental variables. The closeness of pixels is evaluated by a neighborhood distance that is a proxy of the average home-range dispersal distance. The dispersal distance for mammals is, in fact, proven to be proportional to the home-range size \cite{bowman02}. Pixels whose mutual distance is lower than the neighborhood distance are part of the same aggregate.

As for fish and tree species in the MMRS (Figure \ref{fig3}, c and d respectively) \cite{conv11ecomod} determined the aggregate-size spectrum of species by implementing a neutral metacommunity model (NMM) proposed by \cite{muneefish,convertino09}, and further improved by \cite{konar10}. The predicted aggregate-size spectrum match the spectrum calculated using data of species occurrences. The NMM is a stochastic speciation-dispersal model based on the individual per-capita species equivalence assumption. The neutral hypothesis \cite{hubbellbook} holds for the same taxonomic group. An aggregate of fish and tree species is defined as the number of contiguous local communities (a local community is a ``direct tributary area'' \cite{conv11ecomod}) in which a species occurs along the network or according to a Von Neumann neighboring criteria in a 2D domain respectively \cite{conv11ecomod}. For fish and tree specie of the MMRS the aggregates of each species that are assumed equivalent to each other are considered together in determining the aggregate-size spectrum. Thus, the aggregate-size spectrum is representing a metacommunity pattern of species diversity rather than of single metapopulations.

For the African elephant the size and perimeter of elephant aggregates are computed considering the adjacent boxes of the box-counting method (Section \ref{boxc}) at a biologically relevant resolution of grid. We consider as aggregates the boxes whose unitary side length is $38$ km that is the square root of the home range. For the African elephant in the KNP the home range varies from $400$ to $1500$ km$^2$ in the wet (summer) and in the dry (winter) season respectively \cite{grainger05,thomas08,Loarie2009,Marshal10}. The choice of the unitary side length length for definition of the aggregates has a very limited influence on the aggregate-size distribution for scale-free patterns which is also the case of the African elephant \cite{convele}. Unfortunately, for the African elephant we do not have any information about the observed aggregates and the only data available are part of an ongoing project in which a stochastic network-based metapopulation model is implemented \cite{convele} using only occurrences and habitat capacity functions without the requirement of calculating aggregates.

%This is because of the accuracy of the African elephant census performed in the dry season (plane survey), and because there is not an identifiable geomorphic unit associated with the African elephant' movement (like for example a river segment in the case of fishes).

\subsection{Theoretical Construct}\label{theory}

\begin{figure}
\begin{center}
\advance\leftskip-1.5cm
\includegraphics[width=12cm]{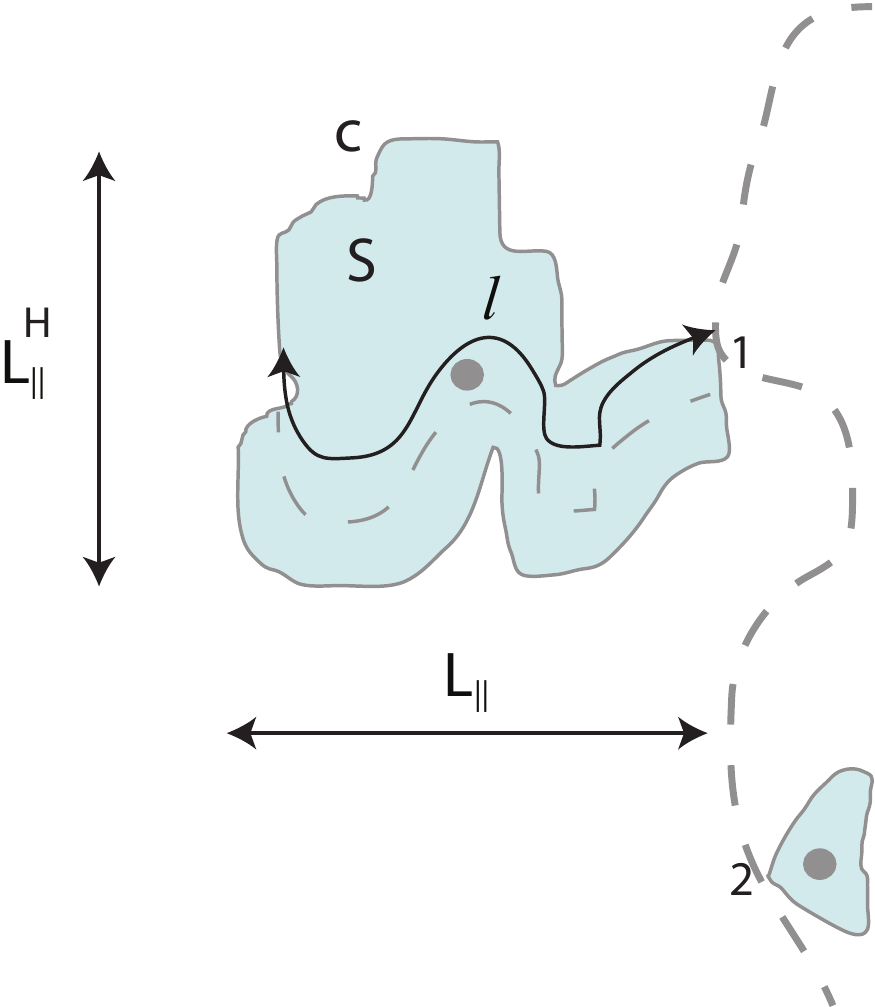}
\caption[h]{} \label{fig1}
\end{center}
\end{figure}

The theoretical characterization of aggregates is based on hypothesis about relationships among aggregate geometrical features. Identical relationships have been formulated for river basins by \cite{Colaiori97}. The allometric ansatz for the size $S$ and the perimeter $C$ of the aggregates are:

\begin{equation}\label{first}
\mbox{} \left\{\begin{array}{ll} S = k_S \; L_{\|}^{D_{\underline{S}}}\\
 \mbox{} C = k_C \; L_{\|}^{D_{\underline{C}}} \\
 \end{array}
\right.,
\end{equation}

where: $L_{\|}$ is the main diameter of aggregates that is a proxy of aggregates' characteristic length and is measured along the principal axis of inertia of aggregates (Figure \ref{fig1} and Figure \ref{fig2}); $L_{\perp}$ is the transversal diameter of aggregates; $D_{\underline{S}} \; = 1+H$ because $S \sim L_{\|}L_{\perp}$ and $L_{\perp} \sim L_{\|}^H$, where $H$ is the Hurst exponent; and $D_{\underline{C}}$ is $d_f$ according to the theory of fractal river basins \cite{fracriv}. The $d_f$ exponent that characterizes the characteristic length of the aggregate characteristic curve, is the fractal dimension of a stream for fractal river networks. A stream that is each rivulet going from any site of the basin to the outlet, is a fractal set with the same fractal dimension along its path. In general, $D_{\underline{S}}$, and $D_{\underline{C}}$ are fractal dimensions related to the morphological structure of aggregates.

The ansatz is to consider that half of the perimeter ($C/2$) scales with a power of one with $l$ that is the mainstream length in river basins (Figure \ref{fig1}). In general $l$ is definable as the length of the aggregate characteristic curve. The aforementioned scaling relationship is verified for river basins. For river basins it was suggested that basin boundaries and mainstream courses are in essence mirror images of each other \cite{Werner91a,Breyer1992,Werner82,ijjasz94,Shankar02}. This assumption generates the second allometric law in Equation \ref{first} irrespectively of the constant. We assume the relationship to hold for any aggregate along a line drawn into the domain on which the aggregates are self-organized (Figure \ref{fig2}) and within any aggregate (Figure \ref{fig1}). The characteristic curve can be the mainstream of a river basin, a rivulet of a subbasin, or any other characteristic curve drawn within the system domain. $S$ can be imagined as the body-mass of a system component in a biological perspective as for river basins in river systems \cite{maritan02}.

From Equation \ref{first} by incorporating the first relationship into the second one, the perimeter-area relationship (PAR) \cite{cheng95} is derived as:

\begin{equation}\label{hack}
C \sim S^{h},
\end{equation}

where $h=d_f/(1+H)=D_{c}/2$ is the Hack's exponent. $h= \frac{D_{\underline{C}}}{D_{\underline{S}}}$ by considering the ansats in Equation \ref{first}. In the ecology literature $D_{c}$ is classically identified as the fractal dimension of an aggregate derived from the PAR. Equation \ref{hack} is commonly known as Hack's law in fluvial geomorphology \cite{Rigon96} where $C$ is the mainstream length and $S$ is the size of a basin. The interchange of the mainstream with the basin boundary is supported by our ansatz and by the empirical evidence of the scaling of the basin perimeter with the mainstream length with a power of one ($C \sim l$) \cite{Werner91a,Breyer1992,Werner82,ijjasz94,Shankar02}. The Hack's law validity is proven in any embedded subbasins within river basins. This shows the self-affinity of river basins and the possibility to extend this law to any system. Here we test the validity of Equation \ref{hack} also for any aggregate of the animate and inanimate systems considered.

The probability density function of the aggregate-size can be universally described by the double-Pareto distribution:

\begin{eqnarray}\label{double}
p(s) =  \frac{  s^{\beta-1} \Theta(s)\Theta(t-s) + s^{-\epsilon-1} \Theta(m-s)\Theta(s-t) } {  \left[  \frac{t^\beta}{\beta} + \frac{m^{-\epsilon}-t^{-\epsilon}}{\epsilon} \right] } &\sim& \nonumber \\  \sim
\left\{ \begin{array}{cc} s^{\beta - 1}  &\mbox{ for  } s < t\\
 s^{- \epsilon -1 } \; f \left(\frac{s}{m} \right) &\mbox{ for  } s > t \\  \end{array}
\right.,
\end{eqnarray}

where, $t$ is the truncation point (``hard truncation'') where a change of scaling can occur, and $m$ is the upper cutoff corresponding to the maximum value of the aggregate-size (Figure \ref{fig3}, e). $f(x)$ is a function such that $f(x) = 1$ if $x\ll 1$ and $f(x) = 0$ if $x\gg 1$. Here $f(x)=\Theta(1-s/m)$. $\beta$ and $\epsilon$ are the scaling exponents of the aggregate-size spectrum. The double-Pareto probability density function (pdf) of the aggregate-size has been widely studied \cite{Reed04}, for example for landslides \cite{stark01,Guzzetti02,stark09}. Here we propose the novel analytical formulation in Equation \ref{double} and we verify if such distribution is reproduced by the box-counting method on data versus model predictions. The probability of exceedence of the aggregate-size is by integration of Equation \ref{double}:

\begin{eqnarray}\label{excee}
P(\geq s)
 = 
  \left\{ \begin{array}{cc} N\; \left[ \frac{t^\beta - s^{\beta}}{\beta}  +  \frac{t^{-\epsilon}  -  m^{-\epsilon}}{\epsilon}  \right] &\mbox{ for  } s < t\\
N \; \frac{s^{- \epsilon} - m^{-\epsilon}}{\epsilon}  &\mbox{ for  } s > t \\
\end{array}
\right.
&\sim& \nonumber \\  \sim
  \left\{ \begin{array}{cc} C _0- s^{\beta} \, C_1 &\mbox{ for  } s < t\\
s^{- \epsilon} \; F\left(\frac{s}{m} \right)  &\mbox{ for  } s > t \\
\end{array}
\right.,
\end{eqnarray}

where, $N =  \left[  \frac{t^\beta}{\beta} + \frac{m^{-\epsilon}-t^{-\epsilon}}{\epsilon} \right]^{-1}$, $C_0$ and $C_1$ are constants, $F$ is a homogeneity function that depends on a characteristic size of aggregates $m \sim S \sim L_{\|}^{1+H}$, and $\epsilon = D_{K}/2$ \cite{Mandelbrot82}. 
$D_{K}$ is the fractal dimension of aggregates. Thus, Equation \ref{excee} is a novel formulation of the Kor\v{c}ak's law \cite{Korcak40} that allows a double scaling regime of the aggregate-size distribution. The distribution is tested against the aggregate spectra predicted by the box-counting and by the models (Section \ref{boxc} and \ref{models} respectively). The fit of the distribution is evaluated by a Maximum Likelihood Estimation (MLE) approach (Section \ref{mle}). The determination of $D_{K}$ is independent from the PAR because it considers only the aggregate-size. In the theory of fractal river-basins \cite{fracriv} a subbasin is a unit of a river basin system subjected to geological and climatological forces. The random variable $s$ is the drainage area of a subbasin in river basin ecosystems. The interaction among subbasins happens along the drainage ridges that divide the runoff among adjacent subbasin hillslopes.

\subsection{Sampling of Aggregate Areas}\label{theory}

For aggregates of systems, we consider the aggregate areas that are distributed in the system along a real or an ideal curved line and along a perfectly straight line in the system domain. The former case is the self-affine case, and the latter case is the self-similar case of aggregates for which $H < 1$ and $H=1$ respectively. The theoretical characterization of the distribution of areas was performed by \cite{conv2007} for subbasins organized along a fractal mainstream, and along a perfectly straight mainstream. Similarly, \cite{LaBarbera01} considered the case of subbasins along a fractal coastline showing indirectly the generality of the theoretical characterization of the area of river basin aggregates. 

The subbasin area contributes to the formation of the drainage area. The drainage area is a cumulative function that is the sum of all subbasins' areas upstream a point which has hydrological and geomorphological implications \cite{fracriv,Dodov05,nardi,Gangodagamage}. The constraint of conservation of the total area \cite{maritan96,fracriv} suggests that the distribution of areas sampled along a given straight line or along a curve where multiple aggregates  occur (i.e. $p_b(s|L_{\|})$, and $p_{ms}(s|L_{\|})$ in \cite{conv2007} respectively), differs from the Kor\v{c}ak's law \cite{Korcak40} for the drainage area $p(s|L_{\|})$ in the scaling exponent. This is supported by empirical evidence for river basins \cite{conv2007}. Indeed if at $i$ sites one collects
the areas $S_i$ and must enforce the constraint $\sum_i
S_i=S_{max}$ (where $S_{max}$ is the total area), the resulting
population is about a different random variable from that leading the exceedence of the drainage area because
the analog areas $S_i$ sampled anywhere do not add to the total area
\cite{fracriv,conv2007}. Hence, this is true for any other system. 

The Hack's exponent is in fact different for the three distributions mentioned above: $h = 1 - \epsilon = 2 - \tau $ for the drainage area; $h = \epsilon = \tau_{ms} - 1$ for the areas along a curve; and $h = 2 - \epsilon = \tau_{b} - 1$ for the areas along a straight line. $\tau$, $\tau_{ms}$, and $\tau_{b}$ are defined in \cite{conv2007} as the scaling exponents of the probability density function of these areas that occur along a curve (self-affine case) or a straight line (self-similar case) (i.e., $1+d_f/(1+H)$, and $2-1/(1+H)$ respectively). In order to obtain the desired distribution of areas the correct $\epsilon$ needs to be introduced in Equation \ref{excee}. 

%
%This is because the subbasin area sampled anywhere within the subbasin is a different random variable, sampled from a different population from those constrained in relation to their possible spatial locations. 

%We show that the probability of exceedence of the drainage area \cite{fracriv} is the Kor\v{c}ak's law \cite{Korcak40} applied to river basins. Hence, subbasins can be considered as aggregates of river basin systems. The analogy of subbasins like aggregates is not new. Considering that the coalescence of system components can be described as an aggregation phenomena along branching trees \cite{Gabrielov99}, \cite{dacosta02} developed a hierarchical cluster system based on Horton-Strahler rules for river networks. 
%
%
% We test the hypothesis plotting the perimeter of the subbasins vs. their area (Figure \ref{fig4}). 

\subsection{Validation of the Theoretical Construct}\label{vali}

Because of the validity of Equation \ref{first}, Equation \ref{hack}, and in analogy with the analytical framework of river basins' drainage area \cite{fracriv}, we assume that the slope of the probability of exceedence of the aggregate-size is $-\epsilon = h - 1$ \cite{fracriv}. The validation of the model for the aggregate-size distribution is tested by comparing: 

\begin{enumerate}

\item the Hack's coefficient derived from the PAR ($h$) versus: (i) $h_K=1-\epsilon = d_f/1+H $ from the Kor\v{c}ak's law for the river basin drainage area (we consider self-affine basins); (ii) $h_K=\epsilon$ for the aggregates of all systems in the self-affine case; and versus (iii) $h_K=2-\epsilon$ for the exact self-similar case of aggregates (for which $H=1$) that is the case of bacteria aggregates. $h_K$ is calculated only from the aggregate-size spectrum;

\item the Hurst coefficient $H$ derived from the scaling relationship $L_{\perp} \sim L_{\|}^H$, versus $H_c=d_f/h-1$ derived from the PAR by assuming an average value of $d_f=1.1$. $H$ is determined only by calculation of the diameters of the aggregates. 

\end{enumerate}

The first validation is to test the relationship between the aggregate-size distribution and the perimeter-area relationship, while the second validation is to test the relationship between the perimeter-area relationship and the allometry relationship of aggregates considering their diameters.

\subsection{Maximum Likelihood Estimation of the aggregate-size Spectrum} \label{mle}

The Maximum Likelihood Estimation (MLE) method here employed was developed in \cite{clauset09} for the selection of the best-fit probability distribution function on data. In this study, power-law (Pareto), the proposed truncated power-law (truncated Pareto-L\'{e}vy) (Equation \ref{excee}), and exponential distributions are tested for the random variable aggregate-size $S$. These distributions are tested on the the aggregate-size calculated by the box-counting and the models explained in Section \ref{models}. The appropriate MLE equation for each distribution is used to derive an exponent with an initial $s_{min}$ parameter set to the minimum value found in the data and model predictions. A best fit dataset is generated with the estimated parameter and a Kolmogorov-Smirnov (KS) test is used to determine the goodness of fit (the KS-D statistic). The KS test is the accepted test for measuring differences between continuous data sets (unbinned data distributions) that are a function of a single variable. 

This difference measure, the KS-D statistic, is defined as the maximum value of the absolute difference between two cumulative distribution functions. We consider the KS-D statistics in a [0, 1] range. The KS-D statistic between two different cumulative distribution functions $P_{N_1}(s)$ and $P_{N_2}(s)$ is defined by $KS-D\;=\; \mbox{max}_{-\infty < s < \infty} \mid P_{N_1}(s) - P_{N_2}(s) \mid$. To determine the best fit value for the $s_{min}$ parameter the calculation is repeated with increasing values for $s_{min}$ taken from the dataset with the value that resulted in the best (lowest) KS-D statistic being retained as the best fit value \cite{clauset09,humphries10}. When fitting a Pareto distribution the method is repeated to derive a best fit value for the $s_{max}$ parameter, so for the Pareto distribution both the $s_{min}$ and $s_{max}$ parameters are fitted in the same way. This method is applied for any scaling regime of the data. The slopes of the exceedence probability are derived using the linear and the jackknife models \cite{Warton06}. The MLE method is used to verify that the proposed Pareto-L\'{e}vy distribution (or commonly called ``double-Pareto'') has the best fit for the observed and predicted aggregate-size spectra.

% \section{Results and Discussion}\label{res}
\section{Results}\label{res}

The Kor\v{c}ak's law for the animate and inanimate systems considered is shown in Figure \ref{fig3} from plot (a) to plot (f) in order of their average aggregate-size that is proportional to the average body-mass of system components. The aggregate-size is calculated by models at different level of complexity, and by image analysis methods as described in Section \ref{models}. The aggregate-size spectrum is tested against the predictions of the box-counting method. In the plots of Figure \ref{fig3} the box-counting relationship is reported with grey squares fitted by a linear regression. The spectra from the box-counting are adjusted by dividing by two the scaling exponent in order to be compared to the Kor\v{c}ak's law spectra that provide the distributions of the aggregate-size with an exponent that is half of the fractal dimension.  
The proposed Pareto-L\'{e}vy distribution (Equation \ref{excee}) has the best fit for the observed and predicted aggregate-size spectra with respect to the other distributions considered by the MLE method (Section \ref{mle}). The KS-D statistic is always higher than 0.87 for this distribution for all the systems considered.

The E. coli bacteria and the Snowy Plover are the systems that exhibit a pure power-law (Pareto distribution) of the aggregate-size for two and three orders of magnitude of the aggregate-size respectively. Fishes and big trees of the MMRS exhibit a truncated power-law distribution of the aggregate-size with finite-size effects (``soft truncation''). The soft truncation is a well-known feature of power-law distributions due to finite-size effects (see \cite{fracriv}). Landslides and the African elephant are the only systems that show a truncated double-Pareto distribution with ``hard truncation''. The hard truncation separate the two scaling regimes of the aggregate-size distribution (Equation \ref{excee}). On the contrary of \cite{stark01} and \cite{Guzzetti02} we are able to reproduce the double-Pareto distribution also for the exceedence probability distribution of landslides. The transition value, from one scaling regime to an other with different exponents of the power-law distribution, is a characteristic value that can be related to the system domain or to biological constraints \cite{sjoberg00,Burroughs01,jovani08,Mashanova10,kefi11}. Double-Pareto size spectra were reported for example for forest fires for which the two scaling regimes were attributed to the two-layer structure of the forest which allows the formation of different kind of fires \cite{Lin09}. However, man-made constraints can exist and influence the spatial distributions of system components, such as fences of the KNP for the elephants \cite{Loarie2009}, and the Petri dish domain for the E. coli \cite{Budrene95}. The influence of strong geometrical constraints on species organization is a very important topic to investigate with process-based models; however, it is outside the scope of this paper. 

In the following we try to discuss some possible origins of the double-Pareto distribution of the aggregate-size for elephants and landslides in the light of our previous comment and because our knowledge of these systems. 
For elephants the social life of male elephants (bulls) and female elephants are very different \cite{Barnes1982,Evans2008}. The females spend their entire lives in tightly knit family groups. These groups are led by the eldest female, or matriarch. Adult males, on the other hand, live mostly a solitary life \cite{Barnes1982}. The spatial distribution of male elephants is more homogeneous than for females and this leads to the one power-law regime of the aggregate-size for male elephants aggregates (black spectrum in Figure \ref{fig3}, e). However, some eldest females are also observed to be solitary especially at the very end of their life. Hence, the aggregate-size distribution of female elephants shows a double power-law regime (orange spectrum in Figure \ref{fig3}, e). Thus, the different power-law structure of the aggregate-size for female and male elephants may be explained by their different social life. This in turn affects the dispersal network and the aggregate-size distribution. For the elephants the variation of $\epsilon$ and $\beta$ is estimated $\pm 0.005$ for a variation of $\pm 10 km$ of the box length of the box-counting that is used to calculate the aggregates. 

For landslides the origin of the double-Pareto distribution has been a matter of debate among geomorphologists. On average small landslides tend to occur much closer to the river network in sites with small hillslope-to-channel distance. On the contrary, large landslides that tend to involve big portions of the hillslopes and their center of mass is further from the network. The fact that the center of mass of most of landslides is always observed futher up on the hillslopes may simply occur due to geomorphological reasons as evidenced in \cite{catani11}. The constraint would be dictated by the dimension of the valley that is expressed by the subbasin ridge to channel distance.  
The double scaling of the landslide-size is also attributed to different triggering mechanisms for example seismic-induced landslides are big and slow phenomena, while storm-induced landslides are small and rapid phenomena. It is also probable that the double-Pareto distribution of the landslide-size is observable because at smaller scales the cohesion forces become prevalent, thus hindering the development of more frequent mass movements; while at larger scales the main resisting mechanical forces are of frictional type only \cite{catani11}. 
The double scaling has also been attributed to undersampling of small landslide events that are difficult to be recorded. The existence of the undersampling effect of small landslides is certain to exist. Nonetheless, independently of any undersampling it was shown that under a given scale the frequency distribution of the landslide-size has a roll-over effect that changes the sign of the first derivative of the Pareto distribution \cite{catani11}. We believe that despite all these suppositions about the origin of the double-Pareto distribution of the landslide-size, the effect of the river network is certainly driving the distribution of landslides.

The collapse test \cite{Stanley99} that verifies the ansatz (Equation \ref{first}) is shown in Figure \ref{fig4} (a). The product $P(\geq s) \; s^{\epsilon}$ has a different constant for each system considered. Thus, we decided to rescale everything to the same constant for better visualization. Two theoretical predictions are validated, namely: the perimeter-area relationship (Equation \ref{hack}); and, the probability distribution of the aggregate-size (Equation \ref{excee}) that is shown to follow a double power-law structure (Figure \ref{fig3}. The collapse test verifies our assumption that the perimeter-size relationship is a more broadly defined Hack's law, and that the power-law distribution of drainage areas is a special case of the Kor\v{c}ak's law for river basins. Values of $H$ from the PAR match the values from direct observations. It is safe to assume, in this context, that $d_f \approx 1$. In all cases of the systems analyzed the theoretical prediction is verified quite well. Overall, the theoretical framework seems consistently verified. We plot the normalized scaling perimeter-area relationship (PAR), $C / C_{max} \sim  (S/S_{max})^{h}$ (Figure \ref{fig4} (b)), because of the large range of perimeters, from a few centimeters of the E. coli bacteria aggregates to the large perimeters of big-tree aggregates of the MMRS.

The first test (Section \ref{vali}) of the the Hack's coefficient derived from the PAR (herein $h$) versus $h_K$ from the Kor\v{c}ak's law is verified. Table \ref{table1} reports the numerical values. Thus, we relate the aggregate-size spectrum with the perimeter-area relationship of the aggregates, while previous studies. For instance \cite{pascualcluster} and \cite{nikora99}) did not find any linkage between the scaling exponents of the two relationships.
The second test ($H_c=H$) appears to be less stringent than the first. It is verified only for the range of variability of the Hack's exponent $0.5 \leq h \leq1$, that is the range commonly observed for river basins \cite{Rigon96}. For $h < 0.5$ and $h > 1$ the Hack's exponent seems well approximated by $H_c=1-\mid d_f/h-1 \mid$. For $h>1$ the edge effect of aggregates is very high, that means that the species are confined in very irregular aggregates. It was demonstrated that the larger the edge-effect determined by the complexity of the aggregate perimeter, the lower the probability of survival for the individual of the species within the aggregate. This is the also the case observed for big landslides, for fishes (supposedly because of the dendritic structure of the river network that create very irregular aggregates), and for E. coli colonies. For $h<0.5$ the compactness of habitat aggregates is very high. For example this is the case observed for the solitary male elephants in the KNP. 
For $ 2 \leq \epsilon +1 \leq 3$ that is the case of big landslides and elephants aggregates a finite mean and infinite variance of the aggregate-size is observed. The general case observed for all the other systems satisfies $\epsilon + 1 < 2$ for which the mean and the variance of the aggregate distribution is infinite. This may lead to the conclusion that the aggregate-size may theoretically increase without an upper limit. However, it is somehow arguable to speculate about mean and variance of aggregate-size to this extent because theoretical studies are required to verify these conclusions.

For the systems analyzed, the Kor\v{c}ak's exponent exhibits a wider range of values than reported in literature \cite{pascualcluster}. We find that the values of the scaling exponent, $\epsilon +1$, is consistent with the range provided by \cite{Juanico05}. For large elephants herds and big landslides (that is for $s \gtrsim 7.0 \times10^3$, and $s \gtrsim 1.0 \times 10^3$ which corresponds to the hard truncation points in Figure \ref{fig3} (e) and (f) respectively) we find a fractal dimension bigger than three. We attribute this singularity to the disproportionate increase of the aggregate length when the unit of measurement (e.g. the box-length of the box-counting method) is decreased. 
Very elongated aggregates for both elephants and landslides are observed. In general, small aggregates tend to be self-similar, while big aggregates tend to be self-affine. The self-affinity (elongation) of aggregates can also be enhanced by geomorphic elements of ecosystems, such as the presence of river networks. This is the case for example of subbasins and fish aggregates. River networks plays a determinant role also in shaping the distribution of riparian trees \cite{Muneepeerakul2008}, and the distribution of elephants \cite{Smit2010}. Both trees and elephants are in fact, water-dependent species and the closeness to water is a fundamental driver of their organization. 
We find different ``fractal domains'' \cite{Grossi01} (or scaling regimes) separated by ``hard'' truncation points of the aggregate-size spectrum. These regimes possibly identify different dynamics of species organization resulting in different aggregate patterns as suggested by \cite{sjoberg00}. With the hard truncation the ``heavy-tailedness'' of the aggregate-size spectrum is less strong than for distributions with finite-size effects. Generally there is a lack of a characteristic aggregate-size in presence of a power-law distribution of the aggregate-size. However, every population of species (or system components) is finite because it is constrained by landscape heterogeneities, anthropic constraints, and/or biological factors. We believe that these factors control together the minimum and the maximum aggregate-size; thus, the distribution of the aggregates \cite{Koponen08}. On the theoretical viewpoint the distribution is scale-free. However, due to the finite size of the population of aggregates we believe it is possible to assign a characteristic scale. We think this is particularly true in the case of ``hard'' truncated power-law distributions, that are in between heavy-tail and log-normal distributions. We underline the importance of a further understanding of the distribution of aggregates for the understanding of system self-organization.

% Assuming the size of the aggregate as the body-mass $M$ we found the scaling relationship $M \sim L{\|}^{d_f/h}$ derived from the PAR and $l \sim L{\|}^{d_f}$, that appears to be species-dependent.

\section{Discussion}\label{discussion}

%The following points are worth mentioning.
%\begin{enumerate}

% box counting 
% theory 
% power law

The study shows that the box-counting provides reliable estimates of the fractal dimension of aggregates using only occurrences of systems components. The box-counting does not capture finite-size effects but it captures hard-truncation points of the aggregate-size spectrum. This is important because different scaling regimes, that are possibly associated with different system dynamics, can be captured by using the box-counting method. Because of the validity of the box-counting that assumes scale-invariance of aggregates, we demonstrate that power-law distributions of the aggregate-size imply fractal patterns as found by other studies \cite{jovani08}. However, the contrary does not hold; scale-invariant patterns are not necessarily realizations of systems with power-law aggregate-size distributions. 
We demonstrate that the box-counting, the perimeter-area relationship, and the Kor\v{c}ak's law provide close estimates of the same fractal dimension. Models of higher complexity provide the smallest estimate of the fractal dimension based on the Kor\v{c}ak's law ($D_{K}$) just using aggregate sizes, while the box-counting provides the largest estimate ($D_b$). The fractal dimension calculated using the PAR ($D_{c}$) is in between $D_{K}$ and $D_b$. Hence, the perimeter-area relationship is possibly the best estimate of the fractal dimension because it considers perimeters and areas of aggregates. Hence, the fractal estimation from the PAR is based on a richer information than other fractal dimensions of the aforementioned methods.

We verify that aggregates can be considered as islands and their perimeter as a curve mirroring the sculpting network in the landscape. We show that the probability of exceedence of the drainage area, and the Hack's law are the the Kor\v{c}ak's law and the perimeter-area relationship (PAR) for river basins respectively. We formulate a probabilistic characterization for animate and inanimate systems extending the fractal theory of river basins to aggregates of any animate and inanimate system. At the system scale aggregates of system components, from bacteria to elephants, are the byproduct of dispersal networks of single individuals, such as for river basins and landslides that are the byproduct of river networks. The Kor\v{c}ak's law, that is the aggregate-size spectrum is verified also for the cumulative drainage area and for the areas of merging subbasins in river basins sampled along a self-similar or a self-affine mainstream. In analogy, mainstreams are for subbasins like Brownian-L\'{e}vy paths of species that disperse in ecosystems.
The ansatz (Equation \ref{first}) is verified by comparing the Hack's exponent, $h$, from the perimeter-area relationship and its estimate derived from the Kor\v{c}ak's law. The Hurst exponent, $H$, from the PAR, is tested against the exponent derived from the allometric relationships between aggregate's diameters. This test is not verified for $h>1$ supposedly because the edge-effect is very high, and for $h<0.5$ because the compactness of aggregates is very high. Both situations are not observed in river basins.

A novel analytical formulation is provided for the probability distribution of the aggregate-size. The analytical formulation is a generalized Kor\v{c}ak's law that describes the double-Pareto and Pareto distributions with finite-size and truncation effects. Double ``fractal regimes'' evidenced by the double-Pareto distribution are possibly signatures of different system dynamics such as it is observed for landslides and elephants. The finite-size effects and the hard truncation in the aggregate-size spectrum are caused by geometrical constraints of the ecosystem (for instance, the maximum extent of the ecosystem that determines an upper limit to the growth of aggregates) or biological constraints. The power-law distribution of the aggregate-size can be a manifestation of the self-organization of species along a network, such as the case of river basins. For the same double-Pareto distribution of the aggregate-size a virtually infinite number of spatial arrangements of aggregates can be generated, likely with different fractal dimensions. Thus, future research is anticipated toward the understanding of the linkage between aggregate-size spectrum and the spatial distribution of aggregates that has relevant consequences for metapopulation dynamics of species, hydrogeomorphological dynamics, and epidemic spreading to name just few examples.

%\end{enumerate}

\section{Conclusions}\label{concl}

%% make this much shorter , move stuff to Discussion

The characterization of systems patterns is crucial as a first step to possibly understand the fundamental drivers of systems processes, and to develop indicators that are capable to predict fluctuations of these patterns. Here we focus on aggregation features of natural animate and inanimate systems and in particular on those that are characterized by a power-law distribution of the aggregate-size. The power-law is manifesting a resilient configuration of the system \cite{kefi11}. Aggregation phenomena are also observed in human systems (for example, cities) and analogies have been drawn between natural and human systems by recent studies \cite{Samaniego08}. We propose the box-counting as a parsimonious null-method for accurately estimating the aggregate-size distribution without the knowledge of any detailed information, rather than system component occurrences, about the systems analyzed. For example we did not use any biological information of the species investigated. The box-counting can be tested against other more ``biologically-complex'' models which provide other complexity measures, such as area, perimeter, and diameters of aggregates. This validation, at least for the cases analyzed, confirms that the occurrences of system components or just the occurrences of aggregates, if available, are enough for the box-counting to predict the aggregate-size distribution reliably.   

The introduced analytical formulation of the aggregate-size distribution can model different Pareto distributions, such as double-Pareto, and Pareto with soft and hard truncations by properly adjusting the distribution parameters. The box-counting does not reproduce the tail of the aggregate-size distribution in presence of finite-size effects. This range of the distribution is very narrow and few system components experience such level of aggregation. However, these systems components are the largest in size; hence, these system components may be vital for the whole systems (for instance when they are the hub of system function). 

% box-counting does not capture the finite size effect 
% exploration of other type of aggregate-size distribution 

Our results show that position, and topological features of any aggregate are determined by global system processes governed by a physical network, a process network or both. The fractal dimension of each aggregate is an estimate of the fractal dimension of the whole pattern of aggregates because aggregates are tightly linked.
We believe that the development of detailed process-based models which result in power-law distributions of the aggregate-size is certainly necessary to verify these conclusions and to test how and which conditions change the aggregate-size distribution from power-law to another type of distribution. However, that is not sufficient if computational and theoretical methodologies for characterizing aggregation patterns, such as the ones here provided, are not available. The aggregate-size spectrum is in fact an important indicator of system form and function. For this motivation methods that capture such organization (for example just by assessing the fractal dimension) and its variation due to endogenous and exogenous changes \cite{jovani08,kefi11} are desired. 

This is also useful for designing animate and inanimate man-made systems with a desired degree of aggregation of system components, multiple levels of aggregation in the same system space dictated by different power-law regimes of the aggregate-size distribution, or time varying aggregation. 

% biologically inspired design of man-made systems with a desired degree of aggregation of system components. control of the fractal regimes for heterogenous aggregation of system component 

\section*{Acknowledgments} 
M.C., I.L., and G.A.K. kindly acknowledge the U.S. Department of Defense - Strategic Environmental Research and Development Program (SERDP) funding, project SI-1699. M.C. acknowledges the funding of project ``Decision and Risk Analysis Applications Environmental Assessment and Supply Chain Risks'' for his research at the Risk and Decision Science Team, Environmental Laboratory, Engineering Research and Development Center, of the US Army Corps of Engineers, in Concord, MA, USA. The authors cordially thank the Eglin Air Force Base personnel, Dr. R.A. Fischer (US Army Corps of Engineers, Engineering Research \& Development Center), US Florida Wildlife Service, and Florida Wildlife Commission, for their help in obtaining information and data of the Snowy Plover. R. Muneepeerakul (ASU) and M. Konar (UIUC) are acknowledged for the fish and tree data of the MMRS. The Kruger National Park (South Africa) is kindly acknowledged for the elephant data. Dr. S. Ferreira (South African National Parks) is acknowledged for overviewing the manuscript and in particular the part about the African elephant in the KNP. The computational resources of the University of Florida High-Performance Computing Center (\url{http://hpc.ufl.edu}) are kindly acknowledged. Permission is granted by the USACE Chief of Engineers to publish this material. The views and opinions expressed in this paper are those of the individual authors and not those of the US Army, or other sponsor organizations.\\

\section*{Author Contributions} 
Conceived and designed the experiments: M.C. F.S.. Performed the experiments: M.C.. Provided support with data: G.A.K., F.C.. Analyzed the data: M.C.. Wrote the paper: M.C., G.A.K.. Revised the paper: G.A.K., I.L.

\newpage

\bibliographystyle{plos2009}
\bibliography{patchtheo3}
%\bibliography{patch2}

\newpage

\clearpage

\thispagestyle{empty}

\section*{Table Captions}

\vskip 1.0cm

{\bf Table 1.}  \textbf{Fractal dimensions, scaling exponents, and validation.} The systems are listed in order of their average aggregate-size which is proportional on average to the body-mass of system components. The box-counting for subbasins, landslides, and E. coli is performed considering the center of mass of the aggregates as point-occurrence patterns. $D_b$, $D_{K}$, $D_{c}$ are the fractal dimensions from the box-counting, the Kor\v{c}ak's law (Eq. \ref{excee}), and the PAR (Eq. \ref{hack}). A double scaling is observed for elephants and landslides. $H$ is derived from the ansatz ($L_{\perp} \sim L_{\|}^H$), $h$ from the PAR. $H_c$ and $h_K$ are compared to $H$ and $h$ for validation of the theory. $h_K$ is derived from the Kor\v{c}ak's law and it is: (i) $1-\epsilon$ from the Kor\v{c}ak's law for the river basin drainage area (we consider self-affine basins); (ii) $\epsilon$ for the aggregates of all the species in the self-affine case ($H<1$); and (iii) $2-\epsilon$ for the self-similar case of aggregates ($H=1$) that is the case of the E. coli. $H_c$ is derived from the PAR and it is: (i) $d_f/h-1$ for $0.5<h<1.0$; and, (ii) $1- \mid d_f/h-1 \mid$ for $h<0.5$ and $h>1$. $\langle L_{\|} \rangle$ is the average aggregate diameter which is a characteristic length of the whole mosaic of aggregates, $S_{max}$ and $C_{max}$ the maximum values for the aggregate area and perimeter. Variation of scaling exponents is estimated $\pm 0.04$. Variabilities of measured exponents are standard errors found by a Maximum Likelihood Estimation method (Section red{mle}) bootstrapping over cases and deriving scaling exponents by the linear and the jackknife models \cite{Warton06}. \\

\newpage

\section*{Figure Captions}

\begin{figure}
\begin{center}
\advance\leftskip-1.5cm
\includegraphics[width=14cm]{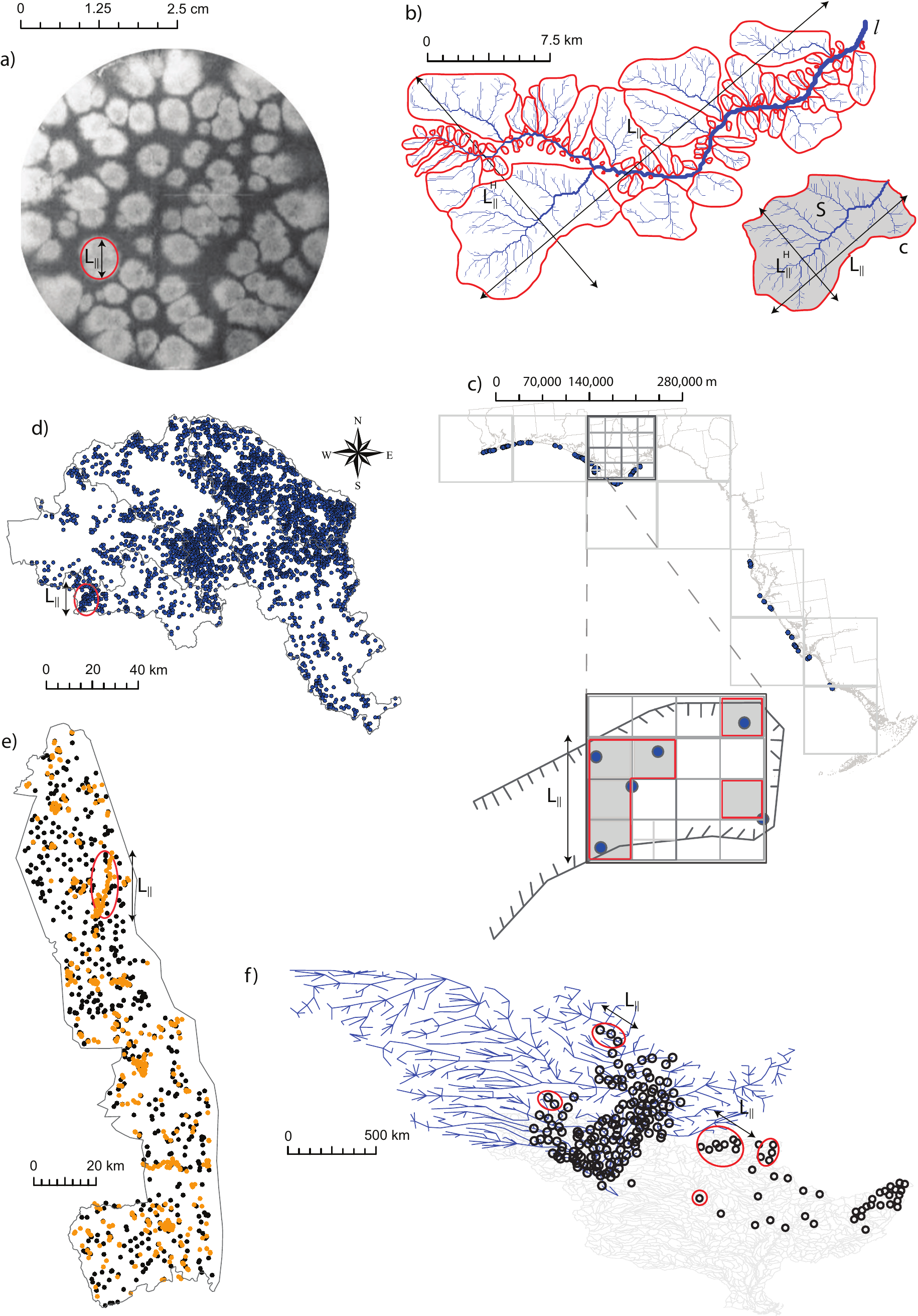}
\caption[h]{} \label{fig2}
\end{center}
\end{figure}

\begin{figure}
\begin{center}
\advance\leftskip-1.5cm
\includegraphics[width=14cm]{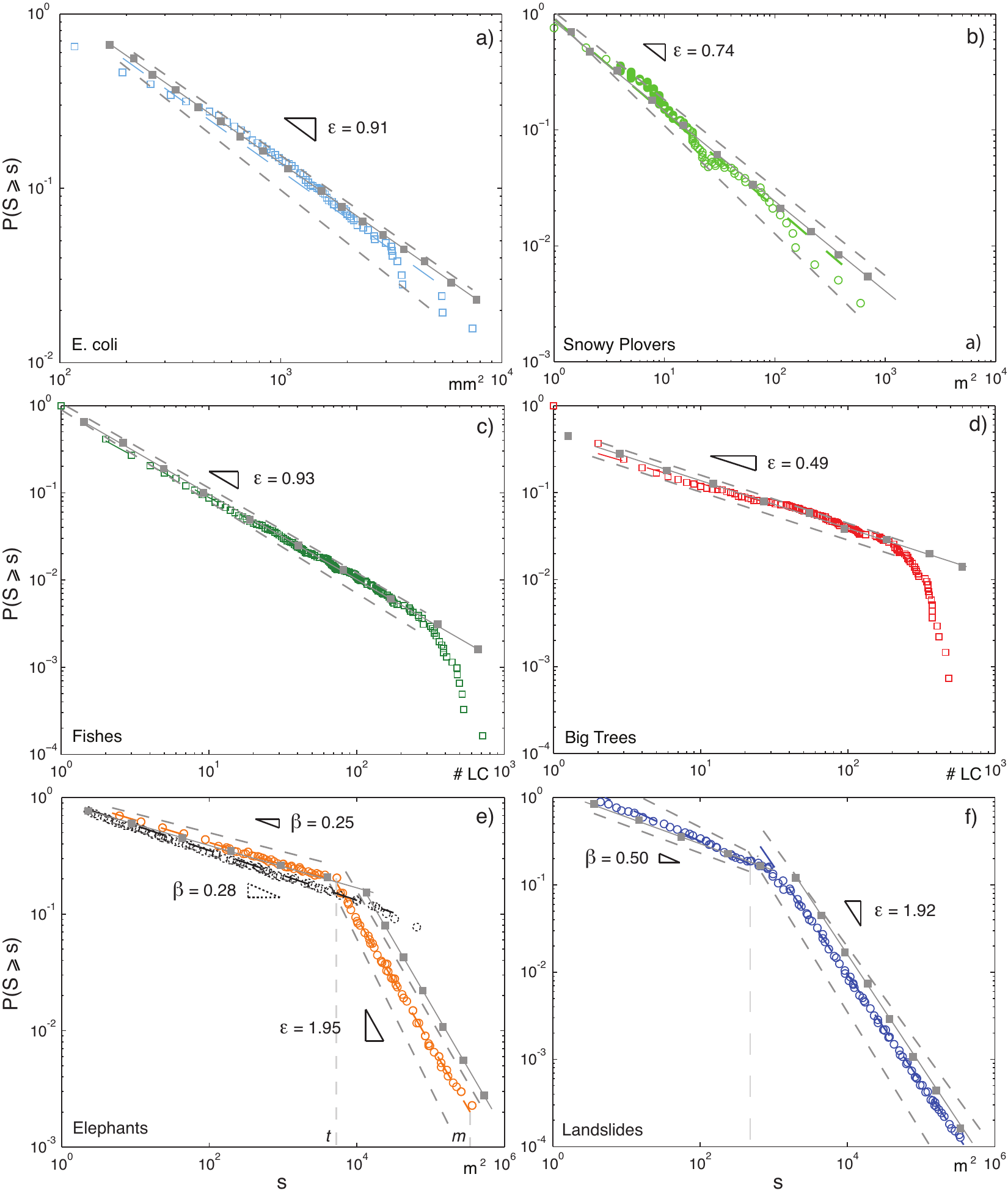}
\caption[h]{} \label{fig3}
\end{center}
\end{figure}

\begin{figure}
\begin{center}
\advance\leftskip-1.5cm
\includegraphics[width=10cm]{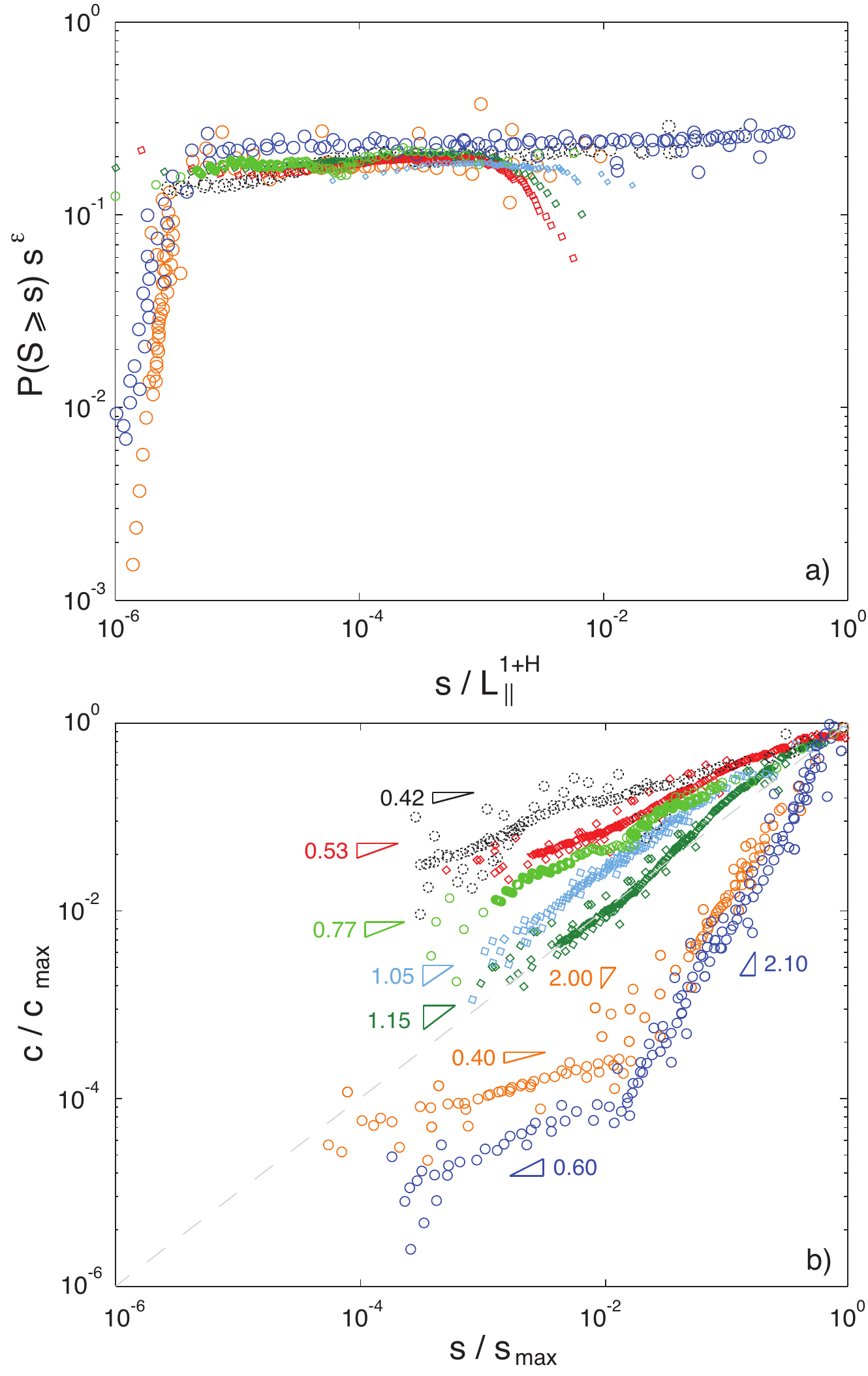}
\caption[h]{} \label{fig4}
\end{center}
\end{figure}

{\bf Figure 1.} \textbf{Schematic representation of the theoretical construct.} An ideal self-similar or self-affine curve is drawn in the system domain where aggregates are self-organized. The curve can be the path followed by species (for instance, a dispersal network similar to a Brownian walk \cite{Benhamou07,cedric11}) or a physical network (such as a river network). Other curves of the same type can be traced within each aggregate. The curve can be imagined as the mainstream of a river. Aggregates are characterized by allometric relationships such as for river basins. $S$ is the aggregate area, $l$ is the length of the curve, $L_{\|}$ and $L_{\|}^H$ are the aggregate diameters. The same quantities are evidenced in Figure \ref{fig2} for river basins. Along the curve it is possible to sample the aggregate areas sequentially (${S_1, S_2, ...}$), or to sample the sum of aggregate areas (${S_1, S_1+S_2, ...}$). This leads to two different probability distributions. The center of masses are represented as grey dots. \\

{\bf Figure 2.} \textbf{Animate and inanimate systems considered in the study.} From (a) to (f) the species are shown in order of the extension of the system domain where they occur. (a) E. coli bacteria colonies (courtesy of \cite{buldyrev03}). (b) Tanaro subbasins identified by their drainage divides in red \cite{conv2007}. (c) Snowy Plover nest occurrences in 2006 along the Florida Gulf coast, and closeup of the box-counting applied to the upper part of the St. Joseph Peninsula State Park \cite{convertinoNATO}. (d) historical Arno landslides from the Synthetic Aperture Radar (SAR) images of the European Remote Sensing spacecraft (the center of mass of landslides is reported) \cite{catani05}. (e) elephant occurrences in 2005 in the Kruger National Park \cite{krugerdata} (plane survey). (f) 100-th most common species of fishes, and of big trees associated to each direct tributary area (DTA, $\sim 3900 \; km^2$) in the Mississippi-Missouri River Basin \cite{muneefish,konar10,conv11ecomod}.\\

{\bf Figure 3.} \textbf{Probability of exceedence of the aggregate-size from model predictions and the box-counting.} The value of the reported scaling exponents $\epsilon$ and $\beta$ is half of the fractal dimension ($D_{K}/2$, Equation \ref{excee}). The probability of exceedence of the aggregate-size is the Kor\v{c}ak's law (Equation \ref{excee}) derived from model predictions. The plots from (a) to (f) are in order of their average aggregate-size which is proportional on average to the body-mass of system components. The aggregate-size unit is reported along on the x-axis. For fishes and big trees of the MMRS the aggregate-size is expressed in ``local-community'' units (LC), where a LC unit is the direct tributary area whose average extension is $3900 \; km^2$. The binned box-counting relationship is reported with grey squared dots. The fractal dimension corresponding to the box-counting method is reported in Table \ref{table1}. The KS-D statistic of the double-Pareto distribution on the aggregate-size from the box-counting is 0.87, 0.90, 0.96, 0.97, 0.93, 0.92 with respect to the other distributions (Section \ref{mle}) for the systems considered from (a) to (f). \\

{\bf Figure 4.} \textbf{Collapse test and perimeter-area relationship.} (a) Intersystems collapse test of the scaling ansatz (Equation \ref{first}). $P(\geq s) \; s^{\epsilon}$ is rescaled to the same constant for all the species. (b) normalized perimeter-area relationship (PAR) (Equation \ref{hack}). The normalized PAR, $C / C_{max} \sim  (S/S_{max})^{h}$, provides a direct estimation of the Hack's exponent $h$.  \\

\newpage
%%% TABLE 
\clearpage

\begin{sidewaystable}
% \begin{table}
\caption{} \centering \label{table1}
\vspace{0.2cm}
%\begin{flushleft} \label{table2}

\begin{tabular}{l|ccccccc}

\hline \hline

% & $SB$ & $L$ & $BT$ & $E$ & $F$ & $EC$ & $SP$  \\ [0.5 ex]

& $\mbox{E. coli}$ & $\mbox{Snowy \; Plover}$  & $\mbox{Fishes}$ &  $\mbox{Big Trees}$ & $\mbox{African Elephant}$ & $\mbox{Landslides}$ &  $\mbox{Subbasins}$ \\ [0.5 ex]

\hline

$D_b$ & $2.10$ & $1.63$ & $2.47$ & $1.33$ & $1.10-4.00$ & $1.23-4.60$ & $1.26$\\

$D_{K}$ & $1.82$ & $1.48$ & $1.86$ & $0.98$ & $0.56-3.90$ & $1.00-3.84$ & $1.04$\\

$D_{c}$ & $2.10$ & $1.54$ & $2.30$ & $1.06$ & $0.80-4.00$ & $1.20-4.20$ & $1.22$\\

\hline
$H$  & $1.00$ & $0.60$ & $0.76$ & $0.85$ & $0.90-0.01$ & $0.80-0.83$ & $0.75$\\

$H_c$  & $0.95$ & $0.43$ & $0.95$ & $1.07$ & $0.80-0.55$ & $0.72-0.53$ & $0.83$\\

\hline 

$h$  & $1.05$ & $0.77$ & $1.15$ & $0.53$ & $1.95-2.00$ & $1.92-2.10$ & $0.61$\\

$h_K$  & $1.09$ & $0.74$ & $0.93$ & $0.49$ & $0.25-0.40$ & $0.50-0.60$ & $0.58$\\

\hline

$\langle L_{\|} \rangle$ & $6.25 \cdot 10^{-5}$ & $2.50 \cdot 10^2$ & $5.00 \cdot 10^4$ & $2.30 \cdot 10^4$ & $5.00 \cdot 10^2$ & $1.50 \cdot 10^3$ & $9.38 \cdot 10^2$\\

$S_{max}$ (m$^2$)  & $6.90 \cdot 10^{-5}$ & $6.20 \cdot 10^2$ & $2.34 \cdot 10^12$ & $1.95 \cdot 10^13$ & $7.00 \cdot 10^{-2}-6.00 \cdot 10^5$ & $6.00 \cdot 10^{-3}-5.50 \cdot 10^5$ & $8.35 \cdot 10^7$\\

$C_{max}$ (m) & $3.00 \cdot 10^{-2}$ & $8.20 \cdot 10$ & $3.40 \cdot 10^6$ & $8.80 \cdot 10^6$ & $1.68 \cdot 10^{-1}-1.55 \cdot 10^3$ & $5.00 \cdot 10^{-2}-1.60 \cdot 10^3$ & $3.60 \cdot 10^4$\\

\hline \hline

% \tableline \tableline
\end{tabular}
\end{sidewaystable}
%\end{flushleft}
% \end{table}

%%% FIGURES 
\newpage
\clearpage \thispagestyle{empty}

\newpage
\clearpage \thispagestyle{empty}

\newpage
\clearpage \thispagestyle{empty}

\newpage
\clearpage \thispagestyle{empty}

\end{document}